\documentclass[draft]{agujournal2019}
\usepackage[T1]{fontenc}
\usepackage[utf8]{inputenc}
\usepackage{textcomp}
\usepackage{url}
\usepackage{lineno}
\usepackage{amsmath}
\usepackage{amssymb}
\usepackage{booktabs}
\usepackage{array}
\usepackage{calc}
\usepackage[inline]{trackchanges}
\usepackage{soul}
\usepackage{placeins}
\usepackage{float}
\usepackage{caption}
\usepackage{multirow}
\usepackage{longtable}
\usepackage{colortbl}
\definecolor{tblhead}{HTML}{D9D9D9}
\definecolor{tblsub}{HTML}{F4F4F4}
\definecolor{tblgrid}{HTML}{808080}

\journalname{Space Weather}

\begin{document}

\title{Systematic Component-Level Characterization of Electricity Transmission Infrastructure for Space Weather Risk Assessment}

\authors{
	Dennies~K.~Bor$^{1,2\dagger}$,
	Edward~J.~Oughton$^{1\ast\dagger}$,
	Evan~A.~Peters$^{1\dagger}$, \\
	Noah~Rivera$^{1}$,
	C.~Trevor~Gaunt$^{3}$,
	Robert~S.~Weigel$^{1,2}$, \\
	Michael~Wiltberger$^{4}$ \\
	\small$^{1}$Department of Geography and Geoinformation Sciences, George Mason University, Fairfax, VA, USA. \\
	\small$^{2}$Space Weather Lab, George Mason University, Fairfax, VA, USA.\\
	\small$^{3}$Dept. of Electrical Engineering, University of Cape Town, Cape Town, South Africa.\\
	\small$^{4}$U.S. National Science Foundation National Center for Atmospheric Research, Boulder, CO, USA. \\
	\small$^\ast$Corresponding author (email: eoughton@gmu.edu).
	\small$^\dagger$Authors contributed equally.
}

\begin{keypoints}
    \item Power grid component data are essential for evaluating geomagnetically induced currents (GICs)
    \item 52,273 components are collected from 1,313 substations, including counts of transformer types
    \item The dataset supports regional GIC screening and first-order GIC estimation during geomagnetic storms, using open-source data
\end{keypoints}

\begin{abstract}
Space weather risk assessment is constrained by the lack of available asset information needed to model geomagnetically induced currents (GICs) in electricity transmission infrastructure. We propose a systematic method that enables risk analysts to collect their own open-source substation data. Using a web browser platform for annotation, we convert OpenStreetMap (OSM) substation locations into high-resolution, component-level mappings of electricity transmission assets. We convert an initial 1,313 high-voltage (\(\geq 230\ \mathrm{kV}\)) substations to 52,273 components using low-altitude, satellite, and Street View imagery accessed through Google Earth, identifying 7,949 transformers. Compared to the OSM baseline, this approach provides detailed insights on voltage levels and substation configurations. We then construct a geospatial GIC network for the Tennessee Valley Authority (TVA) region, comparing May 2024 results with the University of Illinois Urbana-Champaign 150-bus (UIUC150) synthetic network and with measured ground GICs at 13 monitoring devices. The transformer types at unannotated substations and the grounding resistances are unknown, so we sample both across a Monte Carlo ensemble. This gives a median TVA 95th-percentile peak ground GIC of 29.1 A, with a 90\% confidence interval of 23.9--36.1 A. The UIUC150 network yields a 95th-percentile peak ground GIC of 35.8 A under the same forcing, falling within this interval, and the modeled time series broadly capture the temporal morphology of the geomagnetic storm at the monitoring sites. This method shows promise for spatially explicit, screening-level GIC assessment without requiring access to operator data.
\end{abstract}

\section*{Plain language summary}\label{plain-language-summary}
\addcontentsline{toc}{section}{Plain language summary}

Datasets on power grid assets are essential for space weather assessment. Unfortunately, this information is not readily available and rarely shared by network owners. Thankfully, we can collect these data from different types of imagery, such as from satellite and Google Street View imagery. We develop a method to quickly survey substations and collect component information on the assets present, including counts of transformer types, which can then be used to refine vulnerability assessments of geomagnetically induced currents (GICs). We show that the dataset can be used to construct a regional GIC network that broadly captures the temporal morphology of ground currents observed during an actual geomagnetic storm, without requiring access to protected operational parameters from network operators.

\section{Introduction}\label{introduction}

There is grave concern about the threat of severe space weather affecting the critical infrastructure upon which we all rely, particularly for the ongoing operation of the power grid \cite{andreyev_estimating_geomagnetically_2023,cordell_modeling_geomagnetically_2024,dimmock_investigating_trip_2024,smith_sudden_commencements_2024,vandegriff_exploring_localized_2024}. Historical evidence indicates the possible impacts of geomagnetically induced currents (GICs) on extra-high-voltage (EHV) electricity transmission infrastructure. These impacts include power system interruptions and collapse (blackouts), damage to transformers and neutral earthing resistors, and the misoperation of protective relays and reactive power compensation equipment such as capacitor banks and static var compensators. Such consequences motivate risk assessment to support mitigation decisions \cite{akbari_estimation_hot_2023,akbari_thermal_analysis_2023,albert_analysis_long_2022,crack_even_order_2024,espinosa_estimating_geomagnetically_2023,ingham_influence_tectonic_2023,lanabere_analysis_geoelectric_2023,liu_global_observations_2024,marshalko_three_dimensional_2023,nakamura_time_domain_2018,ngwira_occurrence_large_2023,pratscher_modeling_gic_2024,subritzky_assessment_space_2024}. Indeed, EHV transformers are expensive, hard to procure, and difficult to move. These assets are also challenging to repair or replace, highlighting possible supply chain difficulties. Currently there is already a global shortage in EHV transformers, with business-as-usual waiting times reported to be 2--3 years once ordered \cite{moseman_transformer_shortage_2024}. A global spike in demand for these assets, arising from a major space weather event, could cause considerable long-term disruption to society and the economy.

A wide variety of assessments have shown the potential economic impacts of an extreme event up to tens of billions of dollars of daily lost gross domestic product (GDP) \cite{oughton_economic_impact_2018,oughton_quantifying_2017}. The U.S. Government Accountability Office estimates a single GIC-induced blackout could cost up to \$1 billion to the U.S. economy, underscoring the importance of protecting critical infrastructure from space weather impacts \cite{us_government_accountability_office_gao_technology_2018}. Vulnerability to geomagnetic disturbances (GMDs) is one of many hazards capable of degrading transmission assets through mechanical stresses and/or component failure, with the potential to cause large electricity outages \cite{kondrateva_analysis_2020,oughton_stochastic_counterfactual_2019,verschuur_quantifying_climate_2024}. Consequently, there is an urgent need to map, model and pinpoint electricity transmission infrastructure vulnerabilities from natural hazards such as space weather, as identified by successive policy directives, such as via the U.S. National Space Weather Strategy
\cite{sworm_white_paper_2023,sworm_national_space_2019,sworm_national_space_2015}.

One approach over the past decade has been towards developing synthetic power networks which have broadly representative statistical characteristics to support modeling studies \cite{birchfield_statistical_considerations_2017,birchfield_metric_based_2017,li_building_highly_2020,xu_modeling_tuning_2018}. This research has been a step in the right direction, improving the types of data available to assess potential GIC impacts on electricity transmission infrastructure. However, this approach does not enable modelers to understand specific hotspots of risk to their assets and power systems and still leads to considerable uncertainty in identifying where to direct resources and research efforts nationally. Such strategic insight is absolutely required. Thus, our conjecture is that more data can still be collected from emerging geospatial data sources to improve our understanding of electrical transmission infrastructure, such as transformer counts, even if this only provides a partial understanding of the operational parameters in use, such as grounding resistances and transformer winding types. Indeed, authoritative overviews of the methods for calculating GIC emphasize the importance of including appropriate power system characteristics for a modeled circuit, along with magnetic source fields and the present ground conductivity structure \cite{boteler_modeling_2017}. Once general areas of elevated exposure are identified, more specific and detailed modeling and engagement with the relevant utility can be undertaken using their detailed power system data and models. Operational plans can then be implemented to mitigate vulnerabilities, protecting society and the economy \cite{mac_geomagnetically_induced_2023}.

The developments in this paper offer new opportunities for electrical transmission network mapping and validation. By incorporating this approach, researchers can generate more refined datasets describing the types and quantities of important components (e.g., transformers), enhancing the potential for modeling and simulating critical power system infrastructure. Since the required inputs are publicly available, this approach also lets an operator model its position within the wider grid without neighboring utilities having to share protected network data, addressing a long-standing barrier to regional GIC assessment. This paper explores the following four research questions:

\begin{enumerate}
\def\labelenumi{\arabic{enumi}.}
\item
  How does component-level mapping of electricity transmission infrastructure
  enhance our understanding of GIC vulnerability in the U.S. transmission
  network?
\item
  What patterns in component distribution are observable within
  substation assets, and how do these vary by electrical asset
  classification?
\item
  How does this methodology support the systematic expansion and
  refinement of OpenStreetMap data for critical infrastructure mapping?
\item
  How can the annotation dataset be applied to construct a geospatial
  GIC network for regional space weather risk assessment?
\end{enumerate}

In the next section, we undertake a literature review, before articulating a systematic methodology in Section~\ref{sec:methodology} for collecting power network data using open-source imagery, with a Tennessee Valley Authority (TVA) use case in Section~\ref{sec:tva-gic-network-case-study}. In Section~\ref{sec:results}, results are reported before returning in Section~\ref{sec:discussion} to the research questions, discussing these within the broader context of the findings. Conclusions are provided in Section~\ref{sec:conclusion}, along with identification of future research.

\section{Using geospatial data to identify and map electricity
infrastructure assets}\label{using-geospatial-data-to-identify-and-map-electricity-infrastructure-assets}

There is growing interest in using geospatial information to identify and map infrastructure assets. Such information can then be used to improve the accuracy of existing models that simulate both normal operating conditions and disruptions caused by external factors, such as geomagnetic disturbances (GMDs). However, accurate GIC modeling requires appropriate characterization of the power system parameters, including transformer configuration, network topology, and neutral grounding resistance \cite{boteler_modeling_2017}, which are rarely available in open-source datasets, motivating improved component-level data collection of the kind presented in this study. By utilizing geospatial data, it is possible to precisely locate and appraise the characteristics of infrastructure components, such as transformers, powerlines, and reactors for electric grid assets. Although some power operators have highly sophisticated asset tracking and management systems, the long service lives of many of these operating assets mean they were initially deployed many decades ago. Subsequently, the data collection methods reviewed here can also be utilized by network operators to readily map and appraise their own assets, such as for monitoring asset operational condition.

Within power systems, there is a shift towards tracking information on assets by combining geospatial data with remote sensing to monitor key characteristics \cite{fechner_data_2020}. Indeed, remote sensing data have become a valuable resource for energy infrastructure mapping due to its affordability, availability, and accessibility, even in remote regions
\cite{ren_automated_2022}. Ground-based surveying can be expensive and
time-consuming. Although all network lines and substations will have been surveyed prior to construction, not all information may have made its way into a modern asset management platform, such as a Building Information Modeling (BIM) platform, or for use in a more advanced `digital twin' of the ongoing network \cite{ghenai_recent_trends_2022,loggia_electrical_safety_2024}. Thus, remote sensing of assets is an attractive option \cite{ahmed_inspection_identification_2022}, especially to populate new decision-support modeling approaches with necessary data.

Efforts to classify power grid components using remote sensing have advanced substantially, leveraging satellite imagery and public data sources \cite{plaen_towards_open_2024}. One such example employs a combination of satellite imagery and light detection and ranging (LiDAR)-enabled point cloud detection algorithms to extract transformer information from publicly available data \cite{arastounia_automatic_2015}. Previous studies have emphasized the importance of high-resolution imaging in remote sensing, arguing that spatial resolutions between 0.25--1 meter are necessary to capture small but significant infrastructure elements like utility lines and transformers
\cite{jensen_remote_2011}. In cases of more visually prominent features,
resolutions of 1.5--10 meters as provided by SPOT 6/7 and Sentinel-2 are adequate for mapping a global database of transmission components
\cite{kruitwagen_global_2021}. Lower-resolution imagery often results in
misclassification or missed assets.

Machine learning has already enabled automated extraction of transmission line and substation data from remotely sensed imagery, streamlining data acquisition \cite{ren_automated_2022}. Studies also demonstrate how convolutional neural networks can automate infrastructure mapping, providing a scalable solution to detect assets, supporting component-level detection capabilities \cite{bradbury_distributed_2016}. Transfer learning approaches have also been applied to detect transformer locations in distribution networks from publicly available imagery \cite{ali_novel_2020}. However, such automated methods are generally limited to component location rather than to the type- and count-based classifications required for GIC parameterization. Moreover, coordinate-based data in LiDAR point clouds support accurate automated recognition of substation transformers \cite{arastounia_automatic_2015}.

Engaging crowdsourcing platforms like OpenStreetMap (OSM) allows for community-based data collection, filling gaps in remote sensing datasets with local observations. A recent assessment demonstrates the effectiveness of crowdsourcing in mapping for electricity infrastructure, where volunteer-contributed annotation enhanced dataset coverage at the street level \cite{stowell_harmonised_2020}. Similarly, \citeA{ali_generating_2020} demonstrate the use of OSM and automated Python queries to generate open-source datasets for power distribution networks, providing geographical validation of network models at the local scale.

While automated methods can process large datasets quickly, they can struggle to classify equipment accurately in complex or heterogeneous substations \cite{arastounia_automatic_2015}. Examples include distinguishing a three-phase transformer from a bank of three single-phase transformers and distinguishing transformers from reactors. Moreover, the winding configuration, or vector group, of a transformer cannot be determined from imagery, even though it determines whether the transformer provides a path for GIC. For example, a delta-connected winding does not provide a neutral-to-ground path for GIC, whereas a grounded wye-wye winding does. These configurations are visually indistinguishable. Another limitation of this approach is that the route and resistance of underground high- and extra-high-voltage cable circuits cannot be inferred from overhead imagery, so these segments are not characterized. Vertical connectivity may also be obscured, and the open or closed state of circuit breakers cannot be determined from imagery. Manual annotation and expert validation can reduce component misidentification caused by poor segmentation, interpretation, and image resolution \cite{blaschke_object_2010,zanotta_supervised_2018}. However, unobservable electrical attributes require external data or explicit modeling assumptions.

\newpage
\section{Methodology}\label{sec:methodology}

\begin{figure}[H]
    \centering
    \includegraphics[width=1\textwidth]{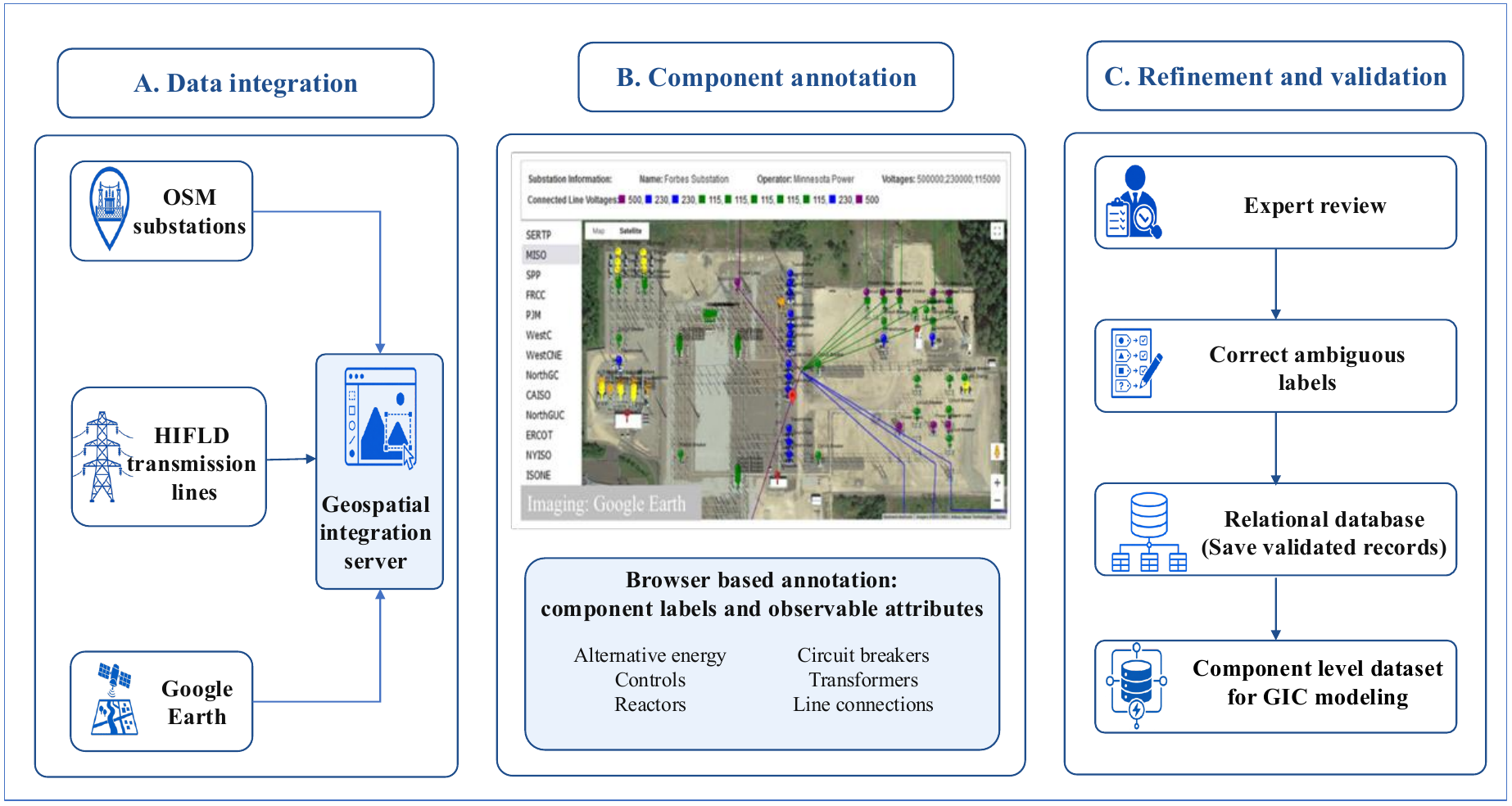}
    \caption{Overview of the methodology for identifying and mapping transmission assets for GIC assessment. (A) Data integration combines OpenStreetMap substations, Homeland Infrastructure Foundation-Level Data (HIFLD) transmission lines, and Google Earth imagery in a geospatial integration server. (B) Component annotation is made using a browser-based interface to label components and their observable attributes. (C) Refinement and validation applies expert review to correct ambiguous labels and stores validated records, yielding the component-level dataset for GIC modeling.}
    \label{fig:methodology}
\end{figure}

We now present a systematic approach for refining OSM data granularity to achieve high-resolution, component-level mapping of electricity substation assets. The methodology is organized into three main phases, as illustrated in Figure~\ref{fig:methodology}. These include Step A Technical Software Design and Data Integration, Step B Initial Asset Annotation Approach and Step C Refinement and Validation. By integrating multiple data sources and a web-based classification interface, we develop a comprehensive component-level dataset designed for improving GIC assessment. Figure~\ref{fig:transmission-results} shows a representative example of annotated substation infrastructure from satellite imagery, illustrating the range of components identified across multiple voltage levels, including transformers, circuit breakers, power lines, and substation controls.

\subsection{Step A - Technical software design and data
integration}\label{sec:methodology-step-a}

Substation location data are sourced from OSM, yielding a dataset of approximately 60,000 generation, distribution, and transmission substations. These substations are categorized into their Federal Energy Regulatory Commission (FERC) grid regions to define their geographical and operational boundaries. Transmission line data are obtained from the Homeland Infrastructure Foundation-Level Data (HIFLD) and geospatially intersected to identify lines connecting to the substations. High-voltage transmission lines and substations rated at 230 kV and above are specifically filtered, as these systems generally have high GIC path flows and are typically more susceptible to GICs than networks operating at lower voltages. Susceptibility also depends on other factors such as ground resistivity.

A custom annotation web interface is then developed using the Google Maps Static Application Programming Interface (API). This tool integrates workflows that enable users to identify and characterize substation components. The web interface is supported by a Python-based server hosted on the Amazon Web Services (AWS) cloud platform and synchronized with a relational database system. For each grid region, substations are listed on the web dashboard, which upon user selection, retrieves Google Earth imagery of the substation. Users can also inspect, annotate, and validate components using Google Street View imagery, if accessible. Only authorized users can update, label, and save data, ensuring secure and controlled data management. Additionally, users can toggle transmission lines intersecting with specific substations, enabling precise mapping and analysis.

This platform allows both technical and non-technical users to access and annotate data collaboratively, enabling updates as annotations are made, and facilitating data entry and interaction with substation assets. The data framework is tailored to substation assets, classified by visual properties such as category and subcategory. Comparative parameters, including geospatial detail, voltage estimations, operational function, and component interconnections, are established to ensure consistent standards across data sources. This standardization allows for efficient data organization and prepares the dataset for annotation and analysis in Geographic JavaScript Object Notation (GeoJSON) or comma-separated values (CSV) formats.

\newpage

\subsection{Step B - Initial asset annotation
approach}\label{sec:methodology-step-b}

\begin{figure}[H]
    \centering
    \includegraphics[width=\textwidth]{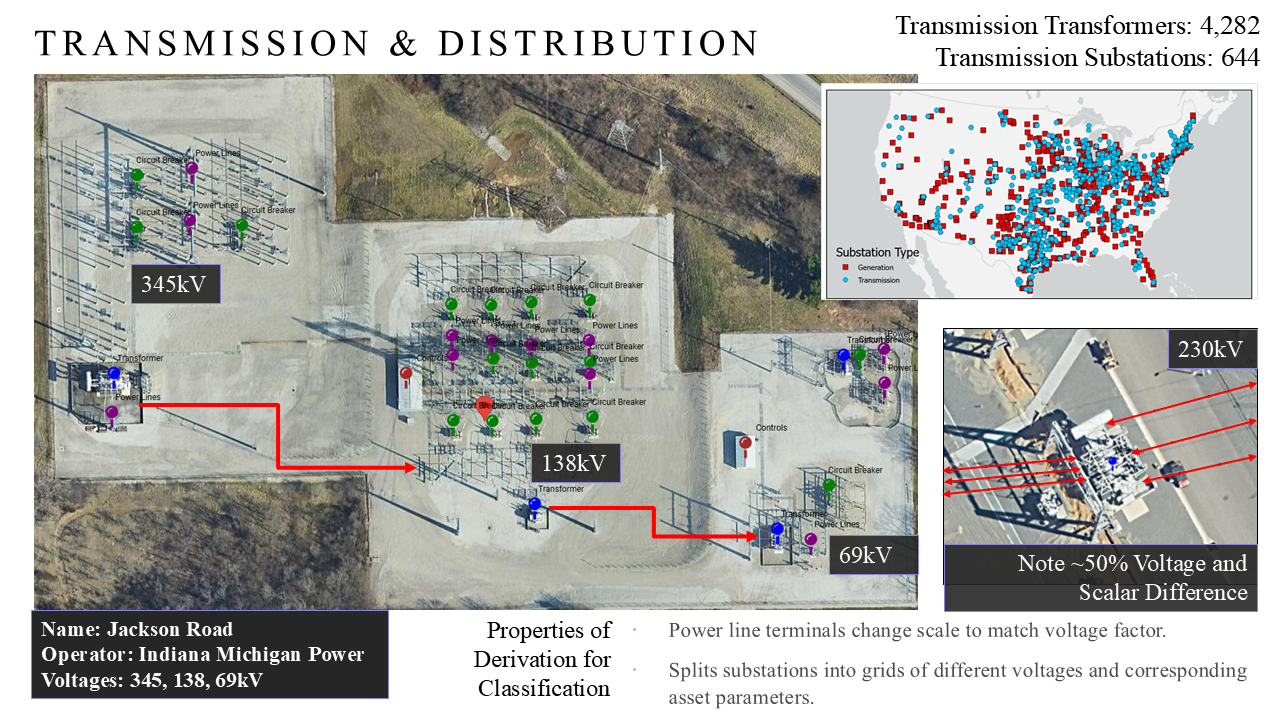}
    \caption{Voltage mapping by spatial configuration. Satellite imagery of Jackson Road substation (Indiana Michigan Power) showing component-level annotation across three voltage sections at 345 kV, 138 kV, and 69 kV. Annotated components include transformers, circuit breakers, power lines, and substation controls. The inset map shows the national distribution of 1,313 annotated substations classified as generation or transmission sites based on the presence and type of transformers identified.}
    \label{fig:transmission-results}
\end{figure}

Individual components within each substation are annotated independently in the web interface, including transformers, transmission line connections, circuit breakers, controls, and alternative energy sources. These marker labels include both the component type as well as additional observable attributes such as phase, utility, and relative voltage category, effectively creating subtypes for enhancing analysis. An initial asset dataset is established, consisting of a single-user annotation for all 1,313 available OSM substations (\(\geq 230\ \mathrm{kV}\)). Each component is visually identified and labeled based on observable features, ensuring that the dataset accurately captures the full range of substation infrastructure components, including transformers, circuit breakers, transmission line connections, substation controls, reactors, isolation assets, and alternative energy sources such as batteries and solar installations, as illustrated in Figure~\ref{fig:transmission-results}. The concentration of annotated substations in the Eastern Interconnection reflects the higher density of high-voltage transmission infrastructure in that region, which is also associated with elevated GIC exposure due to lower ground conductivity, higher magnetic latitude, and proximity to the coast, as demonstrated in \citeA{oughton_major_2025b}.

\subsection{Step C - Refinement and validation}\label{sec:methodology-step-c}

\begin{figure}[H]
    \centering
    \includegraphics[width=\textwidth]{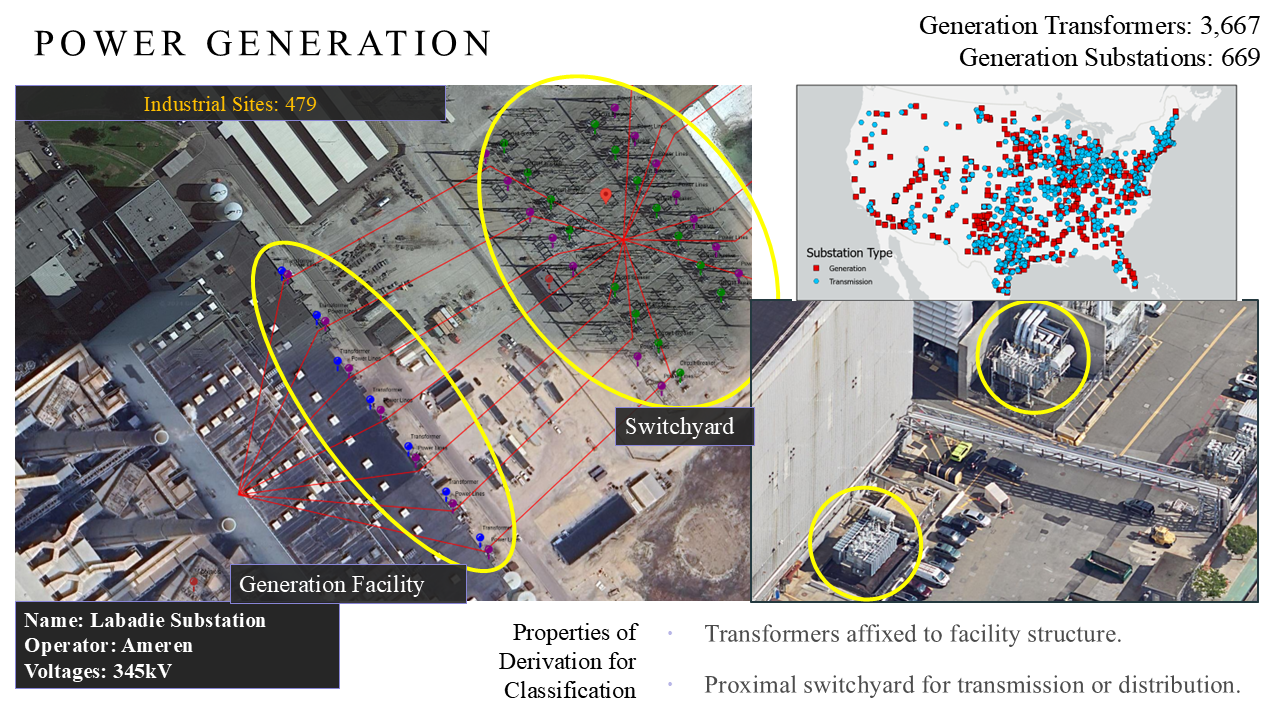}
    \caption{Refining areas of interest for specific component classes, such as generation transformers.}
    \label{fig:generation-results}
\end{figure}

After the initial labeling of assets, further classifications are identified, as illustrated for generation transformers in Figure~\ref{fig:generation-results}. During initial labeling, markers that appeared to be circuit breakers were assigned one of five provisional annotation codes, Types 1--5, based on apparent equipment scale, spatial configuration, connectivity, and contextual OSM voltage information. These codes were working labels rather than standardized circuit-breaker construction types. Expert review determined that markers assigned Type 3 represented disconnect switches and those assigned Type 5 represented lightning arresters. Both were therefore moved to the dataset's isolation-asset category. Types 1, 2, and 4 were retained as overlapping circuit-breaker voltage proxies, with indicative ranges of 345--765 kV, 230--500 kV, and 115--230 kV, respectively. These ranges are not verified nameplate ratings.

Circuit breakers at lower voltages may be possible to identify, although image resolution to enable classification of these smaller assets may be inconsistent (as only urban areas have the highest resolution imagery in Google Earth). For circuit breakers and proximate isolators and transformers, voltage values are assigned using visual cues such as busbar phase spacing and insulator string length, the latter judged from the string's shadow in the imagery and from the number of discs or sheds. These cues are cross-checked for consistency between the connected objects and finalized in the expert validation workshops described below. They are treated as indicative rather than definitive. Insulator length also depends on environmental factors such as pollution level and altitude, which raise the required creepage distance. In addition, rights-of-way are sometimes reinsulated with more hydrophobic materials at the existing spacing, so identical spacing can correspond to different voltage classes. These iterative refinements ensure the collected dataset not only captures the nuances of substation infrastructure but also aligns with operational and modeling requirements for assessing resilience to GMDs and other natural hazards affecting transmission infrastructure, specifically the need for transformer type and count data to parameterize GIC solvers and inform vulnerability assessments. Nevertheless, several biases inherent to the method warrant acknowledgment. Google Earth imagery resolution varies by region, with urban areas typically having higher-resolution imagery than rural areas, which can affect classification consistency. This variability would also pose a significant challenge for any machine learning model trained on this dataset. Moreover, the dataset represents a static snapshot and does not account for assets that have recently been decommissioned or newly commissioned in a rapidly evolving network. Annotator fatigue and subjectivity during classification may also introduce inconsistencies, particularly in complex multi-voltage substations, where component identification requires careful visual interpretation.

Two validation workshops were undertaken with domain experts to support dataset refinement and ensure accuracy. The reviewers were a coauthor with expertise in GIC and power systems and an external power system engineer with experience in transmission protection at electric utilities. Neither was affiliated with the utility examined in the case study, and neither had participated in the annotation. In each workshop, participants were presented with the labeling schema, a sample of completed annotations, and the source imagery. Their role was to review the assigned labels and flag misidentifications, ambiguous cases, and labels inconsistent with the schema. Common categories of correction included transformer-type disambiguation (for example, banks of three single-phase transformers initially labeled as a single three-phase unit), reactor and transformer confusion, and voltage-class assignment at substations spanning multiple voltage levels. Inconsistencies identified during the workshops were corrected before the dataset was finalized, and the labeling protocol was updated where the schema itself was found to be ambiguous. The workshop format is generalizable and could be scaled to broader participation through citizen science initiatives or university training programs, enabling distributed annotation and accelerating coverage of currently unannotated substations.

\subsection{Applying the annotated dataset: TVA GIC network case
study}\label{sec:tva-gic-network-case-study}

The annotation dataset has been applied in two prior studies. \citeA{oughton_major_2025b} used it to construct the contiguous United States reference transmission network for the Coupled Space Weather Impact Model (C-SWIM). The C-SWIM framework links geomagnetic storm drivers to transformer failure probability when exposed to GIC. Furthermore,
\citeA{wilkerson_gic--related_2026} used the same C-SWIM reference network output
against observations during the May 2024 Gannon geomagnetic storm. In this step, we apply the dataset to a regional case study in the TVA. The network construction and GIC solver follow the C-SWIM framework. We deliberately use an open-source network rather than the operator's proprietary model, since the latter is Critical Energy Infrastructure Information (CEII)-protected and unavailable to external researchers. This is the central constraint the method is designed to address, so the resulting network is a first-order approximation rather than a substitute for the operator's validated model. Underground cable segments are not represented, because their routes and resistances cannot be recovered from imagery. The ground GIC measurements were obtained from the TVA monitoring network and the North American Electric Reliability Corporation (NERC) GIC monitoring infrastructure during the Gannon storm \cite{wilkerson_gic--related_2026,wilkerson_2026_20090223}. Together, the annotation dataset and the storm observations support an end-to-end test of the open-source pipeline. Here, we investigate whether a network built from open-source data produces physically reasonable GIC values during the storm when compared with measurements at real monitoring locations and with the community-standard synthetic network. If so, the dataset can support regional GIC screening and first-order GIC estimation during geomagnetic storms without access to CEII-protected operational parameters.

\subsection{Geospatial network
construction}\label{sec:geospatial-network-construction}

\begin{figure}[H]
    \centering
    \includegraphics[width=0.85\textwidth]{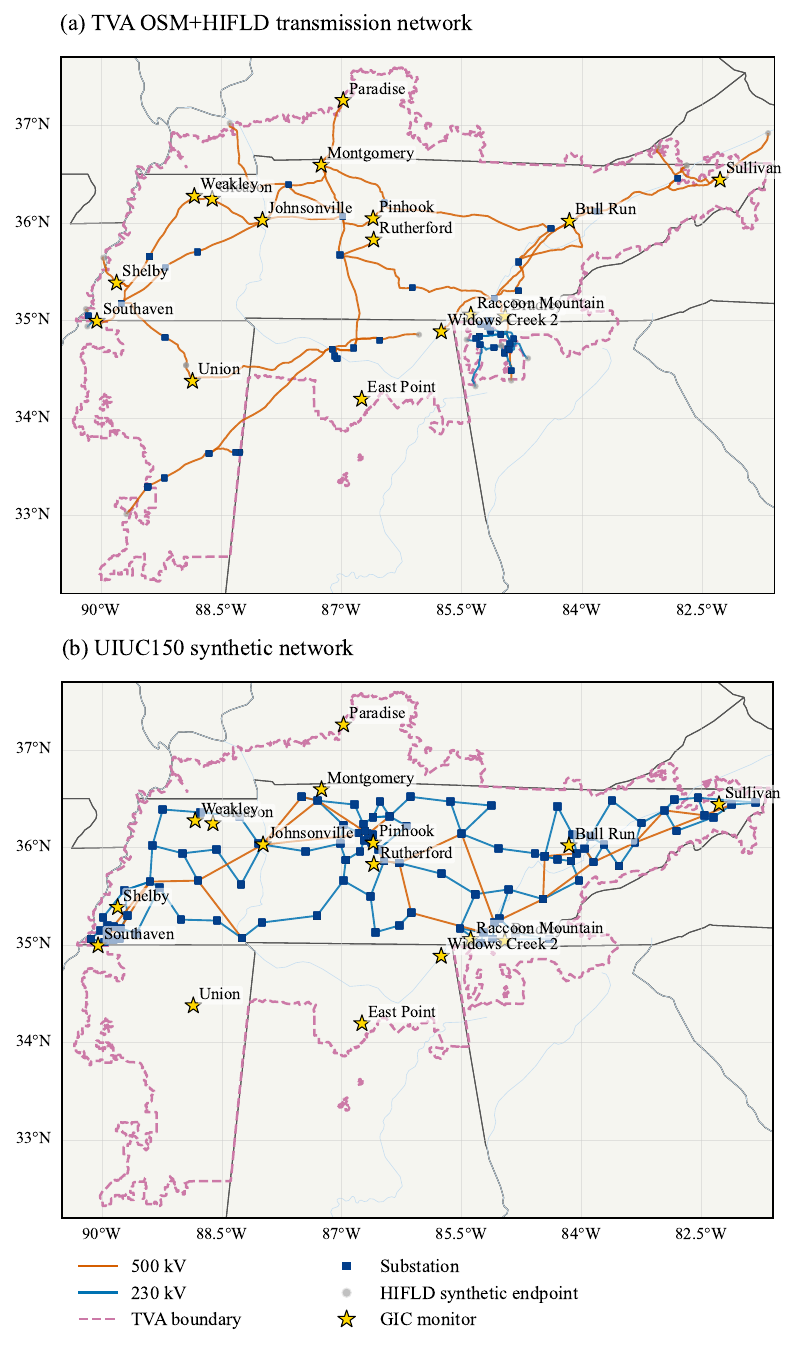}
    \caption{Two-panel map showing the TVA OSM+HIFLD backbone network (top) and the UIUC150 synthetic network (bottom), with the TVA service boundary, monitored substations, and GIC monitoring devices. The UIUC150 network is a synthetic grid statistically matched to regional load and generation, whereas the OSM+HIFLD network is built from open-source geospatial data on substation locations and the geometry of transmission lines.}
    \label{fig:tva-uiuc-backbone}
\end{figure}

The high-voltage transmission infrastructure in the TVA region is derived from two open-source datasets. Substation locations are obtained from OSM and transmission line geometry from the HIFLD database, filtered to operating voltages of 230 kV and above. Substation nodes are linked to transmission line endpoints via a spatial snap distance of 1,500 m, producing a connected backbone network of 715 nodes and 853 edges. Of these, 689 nodes correspond to OSM substations, and 26 are synthetic endpoints introduced where transmission lines do not intersect any OSM substation, even after snapping, typically because the terminating substation is missing from the OSM record. Rather than discarding such lines, which would result in the loss of their contribution to the induced voltage, a synthetic node is inserted at each unmatched line endpoint and treated as a generation substation for parameterization purposes. The largest connected component of this graph is then extracted as the TVA backbone, yielding 95 transmission edges and 84 substations: 60 correspond to OSM locations, and 24 are synthetic. Smaller, disconnected fragments are excluded because they cannot contribute to a coherent network-level GIC computation. Figure~\ref{fig:tva-uiuc-backbone} shows this geospatial network alongside the University of Illinois Urbana-Champaign 150-bus (UIUC150) synthetic power system \cite{birchfield_grid_2017,gegner_methodology_2016}. The synthetic network covers an overlapping geographic footprint within TVA and was designed to reproduce the structural and functional characteristics of real power systems in this region without containing CEII.

\newpage

Transformer counts and types required by the GIC solver are assigned using the annotation dataset described in this study. Of the 60 OSM substations in the TVA backbone, 38 are directly annotated with transformer counts and generation or transmission role classifications derived from satellite imagery inspection. For the 22 unannotated substations, transformer counts are statistically sampled from the distribution of annotated substations. The substation's role determines which transformer types can be assigned. Generation substations draw from generator step-up and grounded wye-delta types, while transmission substations draw from grounded wye-wye, autotransformer, and three-winding types. Each of the 24 synthetic line endpoints is assigned a single generator step-up transformer. Because the transformer types of the unannotated nodes and the substation grounding resistances are not known from open-source data, they are sampled across a 1,000-member Monte Carlo ensemble. Each member draws transformer types from the pools above, according to substation role, and samples each substation grounding resistance from a uniform distribution, using 0.1--1.0 $\Omega$ for all OSM substations, both annotated and unannotated, and 2.0--20.0 $\Omega$ for synthetic endpoints, with roughly 1\% of OSM substations left ungrounded. Line resistances are held fixed at standard reference values, scaling with conductor length using per-kilometer values of 0.010, 0.0141, 0.0283, 0.050, 0.080, and 0.090 $\Omega$/km at 765, 500, 345, 230, 161, and 138 kV, respectively. Transformer winding resistances range from 0.04--0.20 $\Omega$ for grounded windings, while delta-connected windings are treated as open and carry no GIC. These reference resistance values follow the test case of \citeA{horton_test_2012}, a synthetic test case widely used for GIC model validation. Ground GIC results are reported as the ensemble median with a 90\% confidence interval.

\subsection{May 2024 geoelectric field
forcing}\label{sec:may-2024-geoelectric-field-forcing}

The C-SWIM GIC solver is applied to the May 2024 Gannon geomagnetic storm. In order to derive geoelectric fields, geomagnetic field (B) measurements from the International Real-time Magnetic Observatory Network (INTERMAGNET) \cite{kerridge_intermagnet_worldwide_2001} are interpolated across the study region using spherical elementary current systems (SECS) \cite{rigler_interpolating_geomagnetic_2019}, providing B at each magnetotelluric (MT) site. Geoelectric fields (E) are then obtained by convolving the interpolated B with MT transfer functions from the EarthScope MT
\cite{kelbert_first_2019} database, following the methodology described in
\citeA{oughton_major_2025b} and \citeA{lucas_100-year_2020}. Voltage sources along
transmission line segments are computed by integrating the interpolated geoelectric field over the actual HIFLD transmission line geometries, providing a physically realistic forcing that accounts for the spatial variability of the geoelectric field and the true routing of transmission corridors. Ground GIC time series are computed for the 84 identified TVA backbone substations over 4,321 one-minute timesteps spanning the storm period.

\begin{figure}[H]
    \centering
    \includegraphics[width=\textwidth]{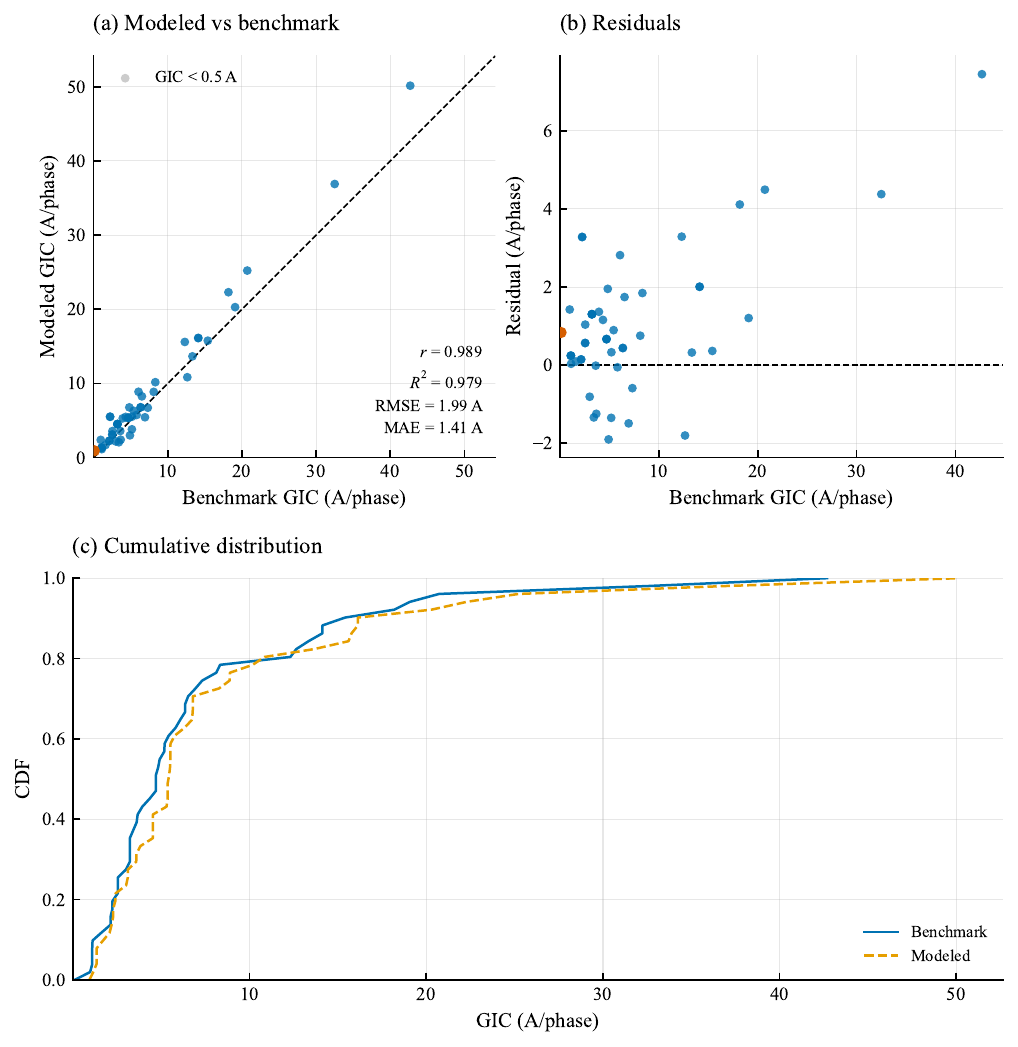}
    \caption{Comparison of modeled and benchmark effective GIC for 52 UIUC150 transformers under a uniform 1 V/km eastward geoelectric field. Panel (a) shows modeled versus benchmark effective GIC per transformer, with a 1:1 reference line. Panel (b) shows residuals as a function of benchmark GIC. Panel (c) shows the cumulative distribution of effective GIC for both the modeled and benchmark datasets, illustrating overall agreement in the distribution of GIC magnitudes across individual transformers.}
    \label{fig:uiuc150-gic-comparison}
\end{figure}

The solver implementation is verified against the published UIUC150 1 V/km benchmark, reproducing the PowerWorld \cite{powerworld_powerworld_2025} effective GIC values across 52 transformers with Pearson $r$ = 0.989, a root-mean-square error of 1.99 A/phase, and a mean absolute error of 1.41 A/phase as shown in Figure~\ref{fig:uiuc150-gic-comparison}.

\section{Results}\label{sec:results}

Here we report the results using the United States as a case study. After eliminating duplicates and limiting distance-based component inclusion per substation, the dataset is refined to a sample of 1,313 outdoor substation assets with 52,273 annotated components, providing a robust foundation for detailed infrastructure analysis and modeling. In the Supporting Information Section S1, we also report the regions of interest, and produce detailed visualizations of the collected data.

Table~\ref{tab:asset-component-counts} presents the composition of substation components across three numerical voltage ranges. Each substation is grouped by the lower bound of its highest annotated component voltage class: \(115 \leq V < 230\) kV, \(230 \leq V < 345\) kV, or \(345 \leq V \leq 765\) kV. These are reporting bins rather than named asset classes. In the bold rows, percentages show the contribution of each component type to all components in that substation group. In the subtype rows, percentages show the subtype composition within the corresponding component type. The final column reports the same measures across all 1,313 substations.

\begin{table}[H]
\centering
\caption{\textbf{Asset and component counts by highest annotated substation voltage range.} The left panel reports total component counts by type and subtype across the 52,273 annotated components. The right panel reports component proportions for 84 substations in the \(115 \leq V < 230\) kV range, 252 in the \(230 \leq V < 345\) kV range, and 977 in the \(345 \leq V \leq 765\) kV range, together with overall proportions across all 1,313 substations.}
\label{tab:asset-component-counts}
\vspace{2mm}

\footnotesize
\setlength{\tabcolsep}{4pt}
\renewcommand{\arraystretch}{0.7}
\arrayrulecolor{tblgrid}

\begin{minipage}[t]{0.29\textwidth}
\vspace{0pt}
\centering
\begin{tabular}[t]{|>{\raggedright\arraybackslash}p{2.35cm}|r|}
\hline
\cellcolor{tblsub}\shortstack[l]{\textbf{Components}\\\vphantom{\textbf{Subtype}}} &
\cellcolor{tblhead}\shortstack[r]{\textbf{52,273}\\\vphantom{\textbf{Subtype}}} \\
\cline{1-2}
\cellcolor{tblsub} & \cellcolor{tblsub} \\
\hline
\cellcolor{tblhead}\textbf{Alt. energy} &
\cellcolor{tblhead}\textbf{2,457} \\
\cellcolor{tblsub}Battery & \cellcolor{tblsub}1,293 \\
\cellcolor{tblsub}Solar & \cellcolor{tblsub}732 \\
\cellcolor{tblsub}Wind & \cellcolor{tblsub}432 \\
\cellcolor{tblhead}\textbf{Circuit breakers} &
\cellcolor{tblhead}\textbf{20,794} \\
\cellcolor{tblsub}CB Type 4 & \cellcolor{tblsub}9,185 \\
\cellcolor{tblsub}CB Type 2 & \cellcolor{tblsub}5,307 \\
\cellcolor{tblsub}CB Type 1 & \cellcolor{tblsub}6,302 \\
\cellcolor{tblhead}\textbf{Controls} &
\cellcolor{tblhead}\textbf{2,931} \\
\cellcolor{tblsub}Industrial & \cellcolor{tblsub}479 \\
\cellcolor{tblsub}Lower voltage & \cellcolor{tblsub}961 \\
\cellcolor{tblsub}Switchgear & \cellcolor{tblsub}424 \\
\cellcolor{tblsub}Upper voltage & \cellcolor{tblsub}1,067 \\
\cellcolor{tblhead}\textbf{Isolation assets} &
\cellcolor{tblhead}\textbf{3,331} \\
\cellcolor{tblsub}Lightning arrester & \cellcolor{tblsub}1,037 \\
\cellcolor{tblsub}Capacitor & \cellcolor{tblsub}1,492 \\
\cellcolor{tblsub}Disconnect & \cellcolor{tblsub}802 \\
\cellcolor{tblhead}\textbf{Power lines} &
\cellcolor{tblhead}\textbf{13,236} \\
\cellcolor{tblsub}External & \cellcolor{tblsub}10,301 \\
\cellcolor{tblsub}Internal & \cellcolor{tblsub}2,935 \\
\cellcolor{tblhead}\textbf{Reactors} &
\cellcolor{tblhead}\textbf{1,575} \\
\cellcolor{tblsub}Air core & \cellcolor{tblsub}1,346 \\
\cellcolor{tblsub}Shunt & \cellcolor{tblsub}229 \\
\cellcolor{tblhead}\textbf{Transformers} &
\cellcolor{tblhead}\textbf{7,949} \\
\cellcolor{tblsub}Gen. 1 phase & \cellcolor{tblsub}916 \\
\cellcolor{tblsub}Gen. 3 phase & \cellcolor{tblsub}2,751 \\
\cellcolor{tblsub}Trans. 1 phase & \cellcolor{tblsub}1,532 \\
\cellcolor{tblsub}Trans. 3 phase & \cellcolor{tblsub}2,750 \\
\hline
\end{tabular}
\end{minipage}
\hfill
\begin{minipage}[t]{0.69\textwidth}
\vspace{0pt}
\centering
\begin{tabular}[t]{|>{\raggedright\arraybackslash}p{2.55cm}|c|c|c|c|}
\hline
\cellcolor{tblsub}\textbf{Component type / subtype} &
\multicolumn{3}{c|}{\cellcolor{tblsub}\textbf{Highest annotated voltage (kV)}} &
\cellcolor{tblhead}\textbf{All 1,313} \\
\cline{2-5}
\cellcolor{tblsub} &
\cellcolor{tblsub}\textbf{115--$<$230} &
\cellcolor{tblsub}\textbf{230--$<$345} &
\cellcolor{tblsub}\textbf{345--765} &
\cellcolor{tblsub}\textbf{Overall} \\
\hline

\rowcolor{tblhead}\textbf{Alt. energy} & \textbf{5.87\%} & \textbf{5.25\%} & \textbf{4.53\%} & \textbf{4.70\%} \\
\rowcolor{tblsub}Battery & 37.17\% & 16.96\% & 61.97\% & 52.63\% \\
\rowcolor{tblsub}Solar & 31.86\% & 30.80\% & 29.43\% & 29.79\% \\
\rowcolor{tblsub}Wind & 30.97\% & 52.23\% & 8.60\% & 17.58\% \\

\rowcolor{tblhead}\textbf{Circuit breakers} & \textbf{28.09\%} & \textbf{40.22\%} & \textbf{40.23\%} & \textbf{39.78\%} \\
\rowcolor{tblsub}CB Type 4 & 100.00\% & 50.90\% & 41.00\% & 44.17\% \\
\rowcolor{tblsub}CB Type 2 & 0.00\% & 49.10\% & 21.54\% & 25.52\% \\
\rowcolor{tblsub}CB Type 1 & 0.00\% & 0.00\% & 37.46\% & 30.31\% \\

\rowcolor{tblhead}\textbf{Controls} & \textbf{7.68\%} & \textbf{5.93\%} & \textbf{5.44\%} & \textbf{5.61\%} \\
\rowcolor{tblsub}Industrial & 8.78\% & 17.00\% & 16.69\% & 16.34\% \\
\rowcolor{tblsub}Lower voltage & 41.89\% & 32.02\% & 32.37\% & 32.79\% \\
\rowcolor{tblsub}Switchgear & 29.05\% & 12.25\% & 14.01\% & 14.47\% \\
\rowcolor{tblsub}Upper voltage & 20.27\% & 38.74\% & 36.93\% & 36.40\% \\

\rowcolor{tblhead}\textbf{Isolation assets} & \textbf{10.33\%} & \textbf{6.07\%} & \textbf{6.25\%} & \textbf{6.37\%} \\
\rowcolor{tblsub}Lightning arrester & 55.28\% & 40.54\% & 27.43\% & 31.13\% \\
\rowcolor{tblsub}Capacitor & 24.62\% & 33.98\% & 48.47\% & 44.79\% \\
\rowcolor{tblsub}Disconnect & 20.10\% & 25.48\% & 24.10\% & 24.08\% \\

\rowcolor{tblhead}\textbf{Power lines} & \textbf{28.66\%} & \textbf{26.49\%} & \textbf{24.93\%} & \textbf{25.32\%} \\
\rowcolor{tblsub}External & 87.50\% & 77.16\% & 77.46\% & 77.83\% \\
\rowcolor{tblsub}Internal & 12.50\% & 22.84\% & 22.54\% & 22.17\% \\

\rowcolor{tblhead}\textbf{Reactors} & \textbf{3.12\%} & \textbf{2.80\%} & \textbf{3.05\%} & \textbf{3.01\%} \\
\rowcolor{tblsub}Air core & 95.00\% & 90.38\% & 84.09\% & 85.46\% \\
\rowcolor{tblsub}Shunt & 5.00\% & 9.62\% & 15.91\% & 14.54\% \\

\rowcolor{tblhead}\textbf{Transformers} & \textbf{16.25\%} & \textbf{13.23\%} & \textbf{15.56\%} & \textbf{15.21\%} \\
\rowcolor{tblsub}Gen. 1 phase & 10.86\% & 5.67\% & 12.57\% & 11.52\% \\
\rowcolor{tblsub}Gen. 3 phase & 27.48\% & 42.29\% & 33.62\% & 34.61\% \\
\rowcolor{tblsub}Trans. 1 phase & 21.41\% & 9.49\% & 20.87\% & 19.27\% \\
\rowcolor{tblsub}Trans. 3 phase & 40.26\% & 42.55\% & 32.94\% & 34.60\% \\
\hline
\end{tabular}

\vspace{1.5mm}
\parbox{\linewidth}{%
\fontsize{8}{9.5}\selectfont
\textit{Abbreviations:} CB (circuit breaker), Gen (generation), Trans (transmission), Alt (alternative). Range lower bounds are inclusive, and upper bounds are exclusive except at 765 kV. Circuit-breaker Types 1, 2, and 4 are relative voltage-proxy categories used in the annotation and post-processing, with indicative ranges of 345--765 kV, 230--500 kV, and 115--230 kV, respectively. They are not standardized breaker construction types or verified nameplate ratings.
}
\end{minipage}

\end{table}
\newpage

Moreover, beyond GIC assessment, the component breakdown in Table~\ref{tab:asset-component-counts} provides a baseline for tracking structural changes in the power grid. The 2,457 alternative-energy components identified document the presence of renewable generation and storage at transmission substations. The annotated dataset could serve as training data for automated machine-learning models to monitor grid evolution at the national scale, with implications for GIC vulnerability assessment, as new transformer configurations associated with renewable integration may alter GIC path flows.

\subsection{GMD screening use case: TVA network under the May 2024 Gannon
storm}\label{sec:gmd-screening-use-case}

\begin{figure}[H]
    \centering
    \includegraphics[width=\textwidth]{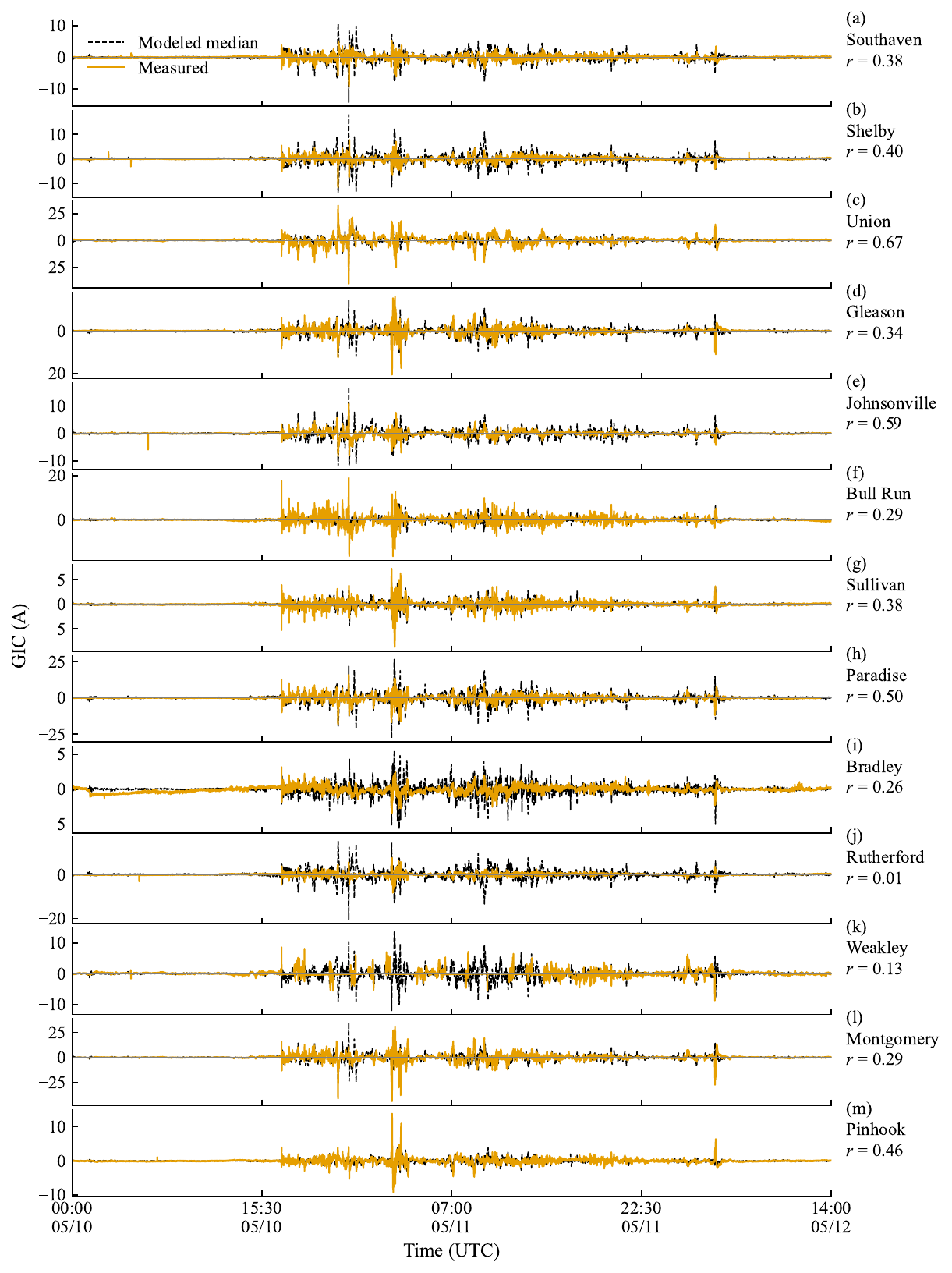}
    \caption{Time series comparison of modeled versus measured ground GIC at 13 TVA monitoring devices during the May 2024 Gannon storm. Each panel corresponds to one monitoring device, with the modeled ground GIC shown as the Monte Carlo ensemble median (dashed black line) and the measured GIC as a solid-colored line. The Pearson correlation coefficient between the modeled median and the measurement, together with the device identifier, is annotated for each site.}
    \label{fig:tva-gic-validation}
\end{figure}

Figure~\ref{fig:tva-gic-validation} compares modeled ground GIC against measurements from 13 TVA monitoring devices. The model captures the temporal morphology of the storm at most monitoring locations, including the primary GIC peak at 22:35 Coordinated Universal Time (UTC) on 10 May 2024. Pearson correlation coefficients between the modeled ensemble median and the measured ground GIC range from $r$ = 0.01--0.67 across the 13 devices, with a median of $r$ = 0.38. Polarity reversals observed at some locations may reflect differences between the current direction used in the model and that used by each GIC monitoring device. Across the Monte Carlo ensemble, the 95th-percentile peak ground GIC has a median of 29.1 A, with a 90\% confidence interval of 23.9--36.1 A, and the maximum at any single substation reaches a median of 66.6 A (90\% confidence interval 44.0--115.9 A). These ground GIC values are reported as the total current flowing through the substation neutral to ground, not as a per-phase quantity, and are therefore approximately three times the corresponding per-phase effective GIC used for comparison with the NERC Transmission Planning Standard TPL-007 threshold of 75 A/phase. For comparison, the measurement at each device is assumed to represent the total neutral-to-ground current at the monitored substation.

Table~\ref{tab:gic-validation} compares the median modeled peak ground GIC across the ensemble against the measured peak at each device. Measured peaks range from 3.4--44.2 A, broadly overlapping the range of the modeled values. The measured peak falls within the modeled 90\% confidence interval at 7 of the 13 sites, and the modeled and measured peaks agree within a factor of two at 10 of the 13 sites.

\begin{table}[H]
\centering
\caption{\textbf{Modeled versus measured peak ground GIC at the 13 TVA monitoring devices during the May 2024 Gannon storm.} The modeled value is the Monte Carlo ensemble median of the per-substation peak, with the 90\% confidence interval (CI) across runs. Measured peak is the maximum absolute GIC recorded at the device. The Pearson correlation coefficient $r$ is computed between the median modeled time series across the ensemble and the measurement. Device locations are given as latitude and longitude in degrees.}
\label{tab:gic-validation}
\vspace{2mm}
\footnotesize
\setlength{\tabcolsep}{3pt}
\renewcommand{\arraystretch}{0.95}
\begin{tabular*}{\linewidth}{@{}l @{\extracolsep{\fill}} l r r c c@{}}
\hline
\textbf{Device} & \textbf{Location (lat, lon)} & \textbf{Measured peak (A)} & \textbf{Modeled median (A)} & \textbf{90\% CI (A)} & \textbf{$r$} \\
\hline
Southaven & 34.99, $-$90.05 & 9.3 & 14.3 & 8.3--30.9 & 0.38 \\
Shelby & 35.39, $-$89.80 & 11.8 & 18.1 & 11.9--40.8 & 0.40 \\
Union & 34.38, $-$88.86 & 40.5 & 28.1 & 22.4--38.2 & 0.67 \\
Gleason & 36.25, $-$88.61 & 20.6 & 15.0 & 8.9--31.0 & 0.34 \\
Johnsonville & 36.03, $-$87.98 & 10.9 & 16.9 & 9.0--41.0 & 0.59 \\
Bull Run & 36.02, $-$84.16 & 19.0 & 8.2 & 4.4--18.5 & 0.29 \\
Sullivan & 36.44, $-$82.28 & 8.7 & 5.0 & 3.4--9.4 & 0.38 \\
Paradise & 37.26, $-$86.98 & 18.5 & 27.8 & 19.9--41.5 & 0.50 \\
Bradley & 35.04, $-$84.96 & 3.4 & 5.7 & 3.4--13.0 & 0.26 \\
Rutherford & 35.83, $-$86.60 & 8.4 & 20.6 & 14.5--35.3 & 0.01 \\
Weakley & 36.28, $-$88.83 & 8.8 & 13.6 & 8.0--28.3 & 0.13 \\
Montgomery & 36.59, $-$87.25 & 44.2 & 34.8 & 22.5--77.7 & 0.29 \\
Pinhook & 36.05, $-$86.61 & 13.9 & 5.6 & 3.0--12.7 & 0.46 \\
\hline
\end{tabular*}
\vspace{1.5mm}
\begin{minipage}{0.9\linewidth}
\fontsize{8}{9.5}\selectfont
\textit{Abbreviations:} lat (latitude), lon (longitude), CI (confidence interval), $r$ (Pearson correlation coefficient).
\end{minipage}
\end{table}

Measured and simulated GIC agree moderately well, though the level of agreement varies across sites. This could reflect five sources of uncertainty in a fully open-source model. First, of the 60 OSM substations in the backbone, 38 have transformer parameters derived from annotations, 22 are sampled statistically, and an additional 24 synthetic nodes are parameterized as generator step-ups. Second, geoelectric fields rely on two layers of interpolation, each with its own error: SECS interpolation of the magnetic field across magnetometer stations to MT sites, followed by interpolation of the derived geoelectric field from MT sites onto transmission lines. Underlying both is the inherent uncertainty in the MT transfer functions themselves, which propagates through the chain. Third, the modeled line resistances are based on standard reference values, and the grounding resistances, which are not supplied by the utility, are sampled from 0.1--1.0 $\Omega$ through the Monte Carlo ensemble. Fourth, OSM and HIFLD records can be outdated and may not capture recent or evolving infrastructure. Fifth, detailed information on the wiring of the monitoring devices and the aggregation of neutral currents at the transformers is unavailable in the open-source record. The comparison therefore assumes that each device measures the total neutral-to-ground current at the corresponding substation, consistent with the modeled quantity. Each of these has a clear path to improvement: extending annotation coverage to additional substations, incorporating resistances supplied by the utility within their service territory while retaining the open-source network for the surrounding system, and complementing OSM and HIFLD entries with operator-supplied records. The agreement reported here, therefore, reflects the lower bound of accuracy that the methodology can deliver.

\begin{figure}[H]
    \centering
    \includegraphics[width=\textwidth]{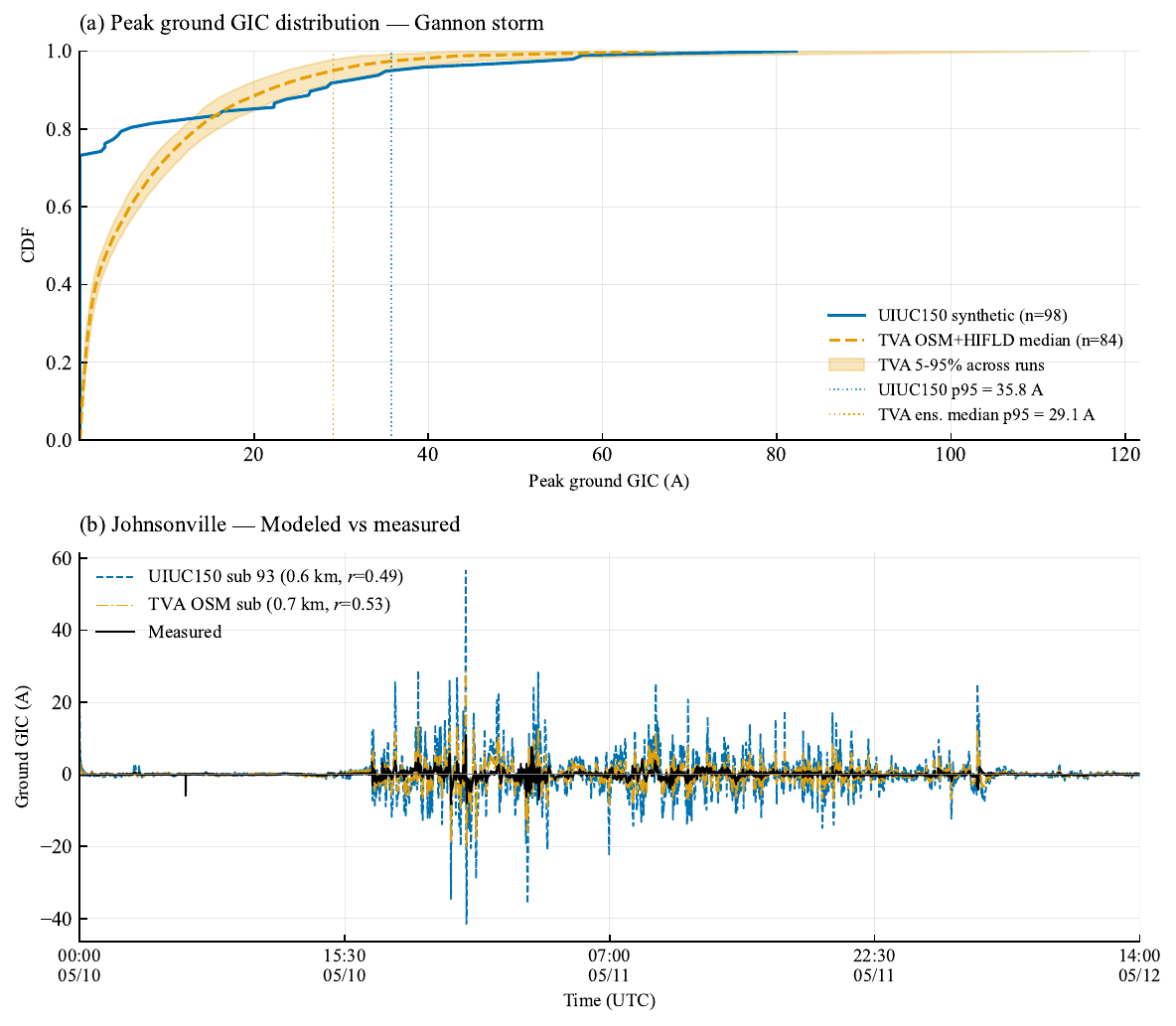}
    \caption{Comparison of modeled ground GIC between the UIUC150 synthetic network and the TVA OSM+HIFLD geospatial network during the May 2024 Gannon storm. Panel (a) shows the cumulative distribution of peak ground GIC. For the TVA OSM+HIFLD network, the median distribution across the Monte Carlo ensemble is shown with a shaded 5th to 95th percentile band, and the vertical dotted line marks the ensemble median 95th percentile (29.1 A). The UIUC150 synthetic network uses fixed published parameters, and its 95th percentile (35.8 A) is shown for comparison. Panel (b) shows modeled ground GIC time series at Johnsonville from both networks alongside TVA measurements, for the nearest UIUC150 substation at 0.6 km and the nearest OSM substation at 0.7 km separation distance.}
    \label{fig:gannon-cdf-johnsonville}
\end{figure}

To place these results in context using the standard synthetic benchmark adopted by the community, the same Gannon-storm geoelectric field is applied to the UIUC150 network, using straight-line bus-to-bus geometry for voltage-source computation. Ground GIC time series are computed for 98 UIUC150 substations across the same 4,321 timesteps. Figure~\ref{fig:gannon-cdf-johnsonville} shows the cumulative distribution of peak ground GIC across both networks, along with a direct time-series comparison at Johnsonville. At Johnsonville, the nearest UIUC150 substation lies 0.6 km from the monitoring device, and the nearest TVA OSM substation lies 0.7 km away. Substations within this separation distance experience broadly similar geoelectric forcing, so their GIC profiles are directly comparable.

\begin{figure}[H]
    \centering
    \includegraphics[width=\textwidth]{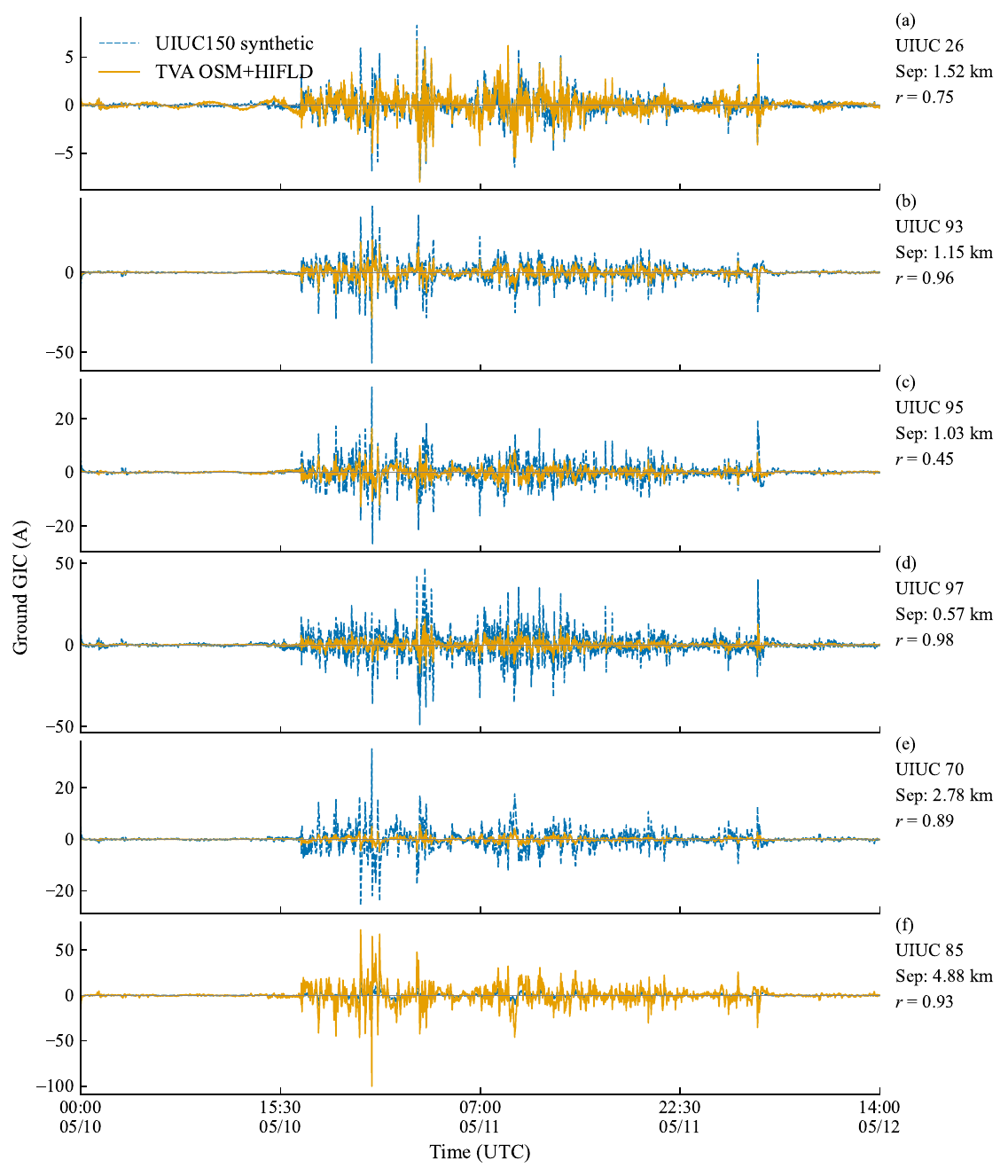}
    \caption{Modeled ground GIC at six matched UIUC150 synthetic and TVA OSM+HIFLD substation pairs during the May 2024 Gannon geomagnetic storm. Each panel corresponds to a matched pair, identified by the UIUC150 substation number and the separation distance (Sep.) between the two substations. The UIUC150-modeled GIC is shown as a dashed blue line, and the TVA OSM+HIFLD-modeled GIC as a solid orange line. Pearson correlation coefficients are computed over the full storm period shown. When the two networks produce negatively correlated time series, the sign of the UIUC150 values is corrected to align with the current reference directions, as noted in the relevant panels.}
    \label{fig:modelled-vs-modelled-gannon}
\end{figure}

The UIUC150 synthetic network, whose grounding and winding resistances are specified in its published dataset, yields a 95th-percentile peak ground GIC of 35.8 A. This value falls within the TVA OSM+HIFLD 90\% confidence interval of 23.9--36.1 A, so the fully parameterized synthetic benchmark is consistent with the open-source network once its parameter uncertainty is accounted for. This is a contextual comparison rather than a validation of either network. Both networks peak simultaneously at 22:35 UTC on 10 May 2024, consistent with the GIC response during the storm being driven primarily by the geoelectric field. Approximately 70\% of UIUC150 substations exhibit zero peak ground GIC during the Gannon storm, reflecting the sparse grounding-path structure of the synthetic network, where generation buses with delta secondary windings lack a neutral grounding path. Figure~\ref{fig:modelled-vs-modelled-gannon} shows modeled time series for six matched substation pairs within 5 km of each other across the two networks. Polarity corrections are applied where the two representations define current direction differently.

\section{Discussion}\label{sec:discussion}

Having outlined the methodology for collecting component-level substation data, and reported results for the case study of the United States, we now return to the research questions.

\emph{How does component-level mapping of electricity transmission infrastructure
enhance our understanding of GIC vulnerability in the U.S. transmission network?}

The component-level mapping in this study provides a more granular dataset for infrastructure analysis. Substations are grouped by their highest voltage, using the highest voltage reported in the OSM substation tags for the baseline and the highest annotated component voltage class for the research dataset. Where the OSM substation voltage is absent, the reported voltage of the connected line is used. Applying a consistent lower boundary of 345 kV to the highest numerical range changes the comparison substantially. The number of substations in the 345--765 kV range decreases from 106 to 92 in the Western region, from 648 to 475 in the Central region, and from 468 to 410 in the Eastern region. The annotation therefore does not identify additional substations in the highest numerical range. Instead, it resolves component configurations and redistributes some substations into lower numerical ranges. This refinement is particularly relevant in regions like the Eastern U.S., where reported ground conductivity structure, latitude, and coastal proximity factors amplify GIC exposure
\cite{kelbert_first_2019}. Accurately characterizing these assets through
component-level annotation enables more informed GMD risk assessments, which in turn support targeted resilience strategies, such as prioritizing transformer upgrades, installing GIC blocking devices at vulnerable neutrals, and implementing operational measures such as network islanding and load switching during storm events \cite{mac_geomagnetically_induced_2023,oughton_assessing_2025a}.

A key advancement of this study is the extensive documentation of circuit breakers, which account for 40\% of all components, or 20,794 in total. Of these circuit breakers, 44\% are classified as Type 4, 26\% as Type 2, and 30\% as Type 1. These annotation categories represent overlapping voltage proxies rather than verified operating voltages or standardized construction types. The distribution reflects the presence of switching equipment across multiple voltage sections within the mapped transmission substations.

The study also sheds significant light on the number and type of present transformers, which make up 15\% of the total components, or 7,949 in total. These are classified into generation and transmission types, with generation transformers comprising 46\% and transmission transformers comprising 54\% of the total.  Transmission substations provide connections among the voltage levels and to load customers. Switching stations, which lack transformers, manage and route electricity across the network. Three-phase transformers dominate the transformer category. The vector group of these three-phase transformers, which governs whether a unit provides a path for GIC, cannot be identified from imagery and requires data supplied by the operator.

In summary, the refined dataset provides a substantial improvement in component-level detail, enabling more nuanced analysis of GIC vulnerabilities across U.S. electricity transmission networks. This work can support the development of targeted resilience measures, ensuring critical assets such as transformers are prioritized and the power system safeguarded against GIC.

\emph{What patterns in component distribution are observable within
substation assets, and how do these vary by electrical asset classification?}

The component-level mapping achieved in this study demonstrates a dramatic increase in granularity, offering a more detailed perspective on substation infrastructure than previously available. The OSM dataset, which represented 1,313 assets, has been expanded to include 52,273 individual components in the new research dataset, an increase in resolution of 3,881\%. This substantial growth in detail enables concentrated analysis of asset distribution and electrical apparatus classification across regions and voltage levels, uncovering patterns that were previously obscured, but are highly important for GIC vulnerability assessment. Indeed, the increased granularity is evident across all regions. In the Western region, resolution expanded from 115 OSM substations to 6,756 components, a 5,775\% increase. The Central region increased by 3,650\% from 689 substations to 25,836 components, while the Eastern region increased by 3,767\% from 509 substations to 19,681 components. At the regional subset level, the California Independent System Operator (CAISO) in the Western subset increased by 8,653\% from 17 substations to 1,488 components. In the Eastern subset, the Southeastern Regional Transmission Planning (SERTP) region increased by 4,468\% from 107 substations to 4,888 components. These improvements in detail provide a way to develop more comprehensive network assessments of transmission infrastructure, reducing model uncertainty, particularly in regions critical to the stability of the grid. This improved resolution directly supports GIC vulnerability assessment, which typically involves computing the effective GIC per transformer under a given geoelectric field scenario and comparing it against thermal capacity thresholds. The NERC Transmission Planning Standard (NERC TPL-007) requires transmission operators to assess transformer thermal risk when effective GIC exceeds 75 A/phase under benchmark geomagnetic disturbance conditions \cite{nerc_tpl_007_2020}. While NERC TPL-007 requires assessment only for transformers, the broader component inventory collected here, including circuit breakers and neutral earthing resistors, supports risk understanding for other primary equipment that GIC can also affect. The transformer count and type classifications provided by this dataset directly reduce parametric uncertainty at each step, enabling more reliable probabilistic assessments of transformer damage under extreme space weather scenarios.

The dataset also reveals mixed patterns in regional voltage averages. The OSM baseline is the mean of the maximum voltage reported for each substation, supplemented by the voltage of the connected line where the substation voltage is absent. The annotated value is the mean component voltage. In the Western region, the regional averages decrease from 453 to 391 kV in the North Connect Grid unconnected region (NGUC), from 467 to 405 kV in CAISO, and from 360 to 337 kV in the West Connect non-enrolled region (WCNE). In the Central region, the Electric Reliability Council of Texas (ERCOT) increases from 324 to 360 kV, while the Midcontinent Independent System Operator (MISO) and Southwest Power Pool (SPP) decrease from 368 to 335 kV and from 334 to 323 kV, respectively. In the Eastern region, ISO New England (ISONE) increases from 344 to 374 kV and the New York Independent System Operator (NYISO) from 317 to 345 kV. PJM Interconnection (PJM) decreases from 394 to 358 kV and SERTP from 476 to 381 kV. These comparisons are descriptive because the OSM and annotated averages summarize different levels of detail, but they show how component annotation resolves the voltage sections within each substation.

\emph{How does this methodology support the systematic expansion
and refinement of OSM data for critical infrastructure mapping?}

The methodology developed in this study establishes a strong foundation for enhancing the accuracy and scope of infrastructure mapping based on OSM data. By leveraging publicly available imagery and integrating it with a web interface, the approach allows for systematic, high-resolution data collection that is accessible, iterative, and capable of refinement over time. The open-source nature of OSM makes it particularly suitable as a baseline resource for collaboration among researchers, utilities, and public agencies, providing a scalable and adaptable framework for critical infrastructure analysis.

This study's detailed component-level data on transformers, transmission lines, and other substation assets demonstrates the potential to substantially refine the granularity and utility of mapping based on OSM's baseline datasets. With 52,273 components mapped across 1,313 substations, the results provide a comprehensive snapshot of current infrastructure, highlighting areas for further refinement and expansion. Continued iterations of this methodology can address gaps in substation layouts, component granularity, and voltage classifications, enabling more detailed mapping of regional and local transmission networks operating at lower voltages. The feedback loop introduced in this research merges refined annotations with coordinate-specific classifications and supplementary information, improving internal dataset accuracy and relevance. This iterative process allows the dataset to evolve over time, adapting to new findings while maintaining alignment with operational priorities and research needs.

A key gap in the existing literature is the lack of labeled component-level data for high-voltage substations that would enable the training and validation of automated methods for GIC-specific asset classification. While automated approaches using OSM and machine learning have advanced substantially, they have primarily focused on distribution networks and general component location rather than on the type- and count-based classifications required for GIC modeling. The 1,313 annotated substations described here constitute a labeled training dataset that directly addresses this gap, supporting future development of automated inference models for national-scale transformer count and role classification from satellite imagery. Satellite imagery used in this study can also help validate OSM-derived substation locations, thereby improving the geospatial realism of constructed grids. Indeed, we acknowledge that power system operators maintain authoritative asset data and validated power system models within their asset management and geographic information systems. Therefore, the resource presented here is best positioned to support academic studies of vulnerability to extreme events and socioeconomic impacts and to facilitate closer partnerships between utilities and academics rather than replace operator data.

Although not directly relevant for GIC assessment, many substations at lower voltages remain unaccounted for or may be under-analyzed. The techniques developed in this study offer a practical means to address these gaps, for outdoor assets visible from space, enhancing our understanding of network structure at lower voltages. This is indirectly relevant when a power outage occurs at a substation operating at higher voltage (e.g., from GIC), as when cascading failure occurs it would be useful to know which areas are likely to be affected at substations operating at lower voltages.

\emph{How can the annotation dataset be applied to construct a
geospatial GIC network for regional space weather risk assessment?}

The transformer counts and types collected in this study provide key structural inputs for building a physics-based GIC network without access to CEII-protected data. Of the 60 OSM substations found in the TVA backbone, 38 are directly parameterized from annotated counts and role classifications. The remaining 22 are sampled from the annotated distribution, with the 24 synthetic nodes assigned generator step-up transformers. Transformer types and grounding resistances are sampled across a 1,000-member Monte Carlo ensemble to propagate the uncertainty in these open-source parameters. The in-service status of circuit breakers and switches cannot be determined from imagery and is not provided by the operator, so the network assumes a nominal fully connected topology. Switching states alter GIC pathways during an event, and capturing them would require operator supervisory control and data acquisition (SCADA) data or an ensemble over plausible network configurations. The resulting network, built entirely from open-source OSM and HIFLD data, produces a 95th-percentile peak ground GIC with an ensemble median of 29.1 A and a 90\% confidence interval of 23.9--36.1 A under the Gannon storm. The UIUC150 synthetic network, whose parameters are fully specified in its dataset, yields a 95th-percentile peak ground GIC of 35.8 A under the same forcing, which falls within this interval. This indicates that, for tail-risk screening, the choice between an OSM-derived network and a synthetic network matched to regional load and generation yields comparable results. Modeled time series at the 13 TVA monitoring locations also capture the storm morphology. Together, these results indicate that an open-source pipeline can produce GIC magnitudes and patterns at the regional scale that are useful for risk assessment, motivating a closer look at where this contribution sits within the broader space weather risk framework.

Space weather risk assessments are commonly structured around the likelihood, exposure, and vulnerability framework \cite{baldwin_vulnerability_2023,meiler_uncertainties_2023}. This study advances the exposure component by characterizing the substation assets susceptible to GIC, with component-level transformer counts and generation or transmission role classifications derived from satellite imagery. The power grid infrastructure, its substations, transformers, and transmission line connections, is the mechanism through which geomagnetic disturbances are translated into induced currents, so accurate asset characterization is a prerequisite for moving beyond simple screening toward a full vulnerability assessment.

This is a meaningful step beyond the simpler approaches historically used for regional GIC screening, including \(\mathrm{d}B/\mathrm{d}t\) as a direct proxy \cite{viljanen_time_2001} and \(E\) divided by line resistance, computed outside any network model \cite{lucas_100-year_2020}. Both ignore network topology, transformer grounding paths, and the integration of the geoelectric field through a connected circuit. The pipeline demonstrated here includes all three, and the GIC magnitudes computed in Section~\ref{sec:gmd-screening-use-case} from this annotated network feed directly into transformer thermal fragility functions, which relate effective GIC per transformer to the probability of thermal failure and enable probabilistic estimation of transformer damage under a given storm scenario, following the framework in \citeA{oughton_major_2025b}.

A practical consequence is that utilities can use the open-source network as a regional baseline and refine it in their own service territory with measured resistances and verified transformer parameters, while continuing to rely on the open-source network for the surrounding system. This addresses a long-standing problem: utilities cannot share data from neighboring utilities while still needing to model their position within the wider grid. The same logic extends to other regions and countries, since the pipeline's inputs are publicly available and the annotation method itself does not depend on local data-sharing agreements.

The 1,313 annotated substations across the continental United States also constitute a labeled training set. Future work can use these annotations to train computer vision models that infer transformer counts and types for the roughly 50,000 unannotated U.S. substations from satellite imagery, OSM geometry, and network topology, enabling national-scale GIC risk assessment with reduced parametric uncertainty. The pipeline can also estimate GIC during a geomagnetic storm, since the magnetic field data, magnetotelluric transfer functions, and transmission line geometry it requires are all publicly available. Such estimates are first-order, since computing actual GIC flows would require knowledge of the operator's network configuration, transformer vector groups, and measured grounding resistances, none of which are observable from open-source data.

\section{Conclusion}\label{sec:conclusion}

This study set out to provide a systematic and generalizable methodology for collecting component-level electricity transmission data to enable improved GIC assessment. Here, we use the U.S. as a case study, substantially enhancing the granularity and accuracy of electricity transmission assets compared to what is currently available. For example, by focusing on improving substation asset data for modeling and resilience planning, we leveraged publicly available imagery and annotations integrated with OSM data to create a high-resolution, component-level dataset. Spanning 52,273 identifiable components, this dataset categorizes critical components important for GIC vulnerability assessment, such as transformers, circuit breakers, and transmission lines, and offers a large set of detailed substation configurations. The resulting methodology not only improves data accuracy but also provides a robust foundation for scalable, collaborative mapping efforts. This framework supports the development of infrastructure models essential for analyzing vulnerabilities, including those associated with GMDs and extreme meteorological disturbances, while advancing the broader goal of enhancing infrastructure resilience. The method can be readily applied to other countries.

The dataset is further demonstrated as a direct input to physics-based GIC modeling in Sections~\ref{sec:geospatial-network-construction} and \ref{sec:gmd-screening-use-case}, where transformer counts and type classifications derived from satellite imagery are used to parameterize a geospatial GIC network for the TVA region. Driven by the May 2024 Gannon geomagnetic storm, the open-source network produces a 95th-percentile peak ground GIC with an ensemble median of 29.1 A (90\% confidence interval 23.9--36.1 A). The fully parameterized UIUC150 synthetic network yields a 95th-percentile peak ground GIC of 35.8 A under the same forcing, which falls within this interval, and the modeled time series at 13 TVA monitoring devices capture the temporal morphology of the storm. The results show that the annotated dataset supports regional GIC screening and scenario analysis without requiring access to CEII-protected operational parameters.

The dataset also provides a regional baseline that utilities can refine in their own service territory by substituting measured resistances and verified transformer parameters, while continuing to use the open-source network for the surrounding system. This addresses a long-standing problem in which operators cannot share data from neighboring utilities, yet still need to model their position within the wider grid. The methodology and the resulting dataset also have applications beyond GIC. Any hazard whose impact depends on component-level network structure, including extreme weather, wildfire, and physical attack scenarios, can use the same workflow. Synthetic networks built by statistical matching of load and generation distributions cannot support this kind of analysis because they do not reflect the actual grid. The annotated dataset does, which makes it useful as an input to broader infrastructure risk modeling.

The quantitative findings highlight the improvements achieved through this approach. The mapping expanded from 1,313 substation-level records to 52,273 component-level records. At the regional level, the mapping expanded from 115 OSM substations to 6,756 components in the Western region, from 689 to 25,836 in the Central region, and from 509 to 19,681 in the Eastern region. These changes correspond to increases of 5,775\%, 3,650\%, and 3,767\%, respectively. Applying consistent numerical voltage boundaries also corrects the earlier classification. Substations in the 345--765 kV range decrease from 106 to 92 in the Western region, from 648 to 475 in the Central region, and from 468 to 410 in the Eastern region after component annotation. Together, the component inventory, voltage sections, line capacities, and possible connection points provide a stronger basis for infrastructure parameterization, GIC risk assessment, and uncertainty analysis.

Despite these advancements, limitations were identified that indicate opportunities for future research. Individual transmission lines are heavily reliant on imagery of the highest resolution. Due to distortion and lack of clarity, tracing the exact pathways of internal substation conductors was avoided, while HIFLD transmission line geometries were used for the network connections. The absence of burial tags for underground conduits constrained the accuracy of mapping in urban regions like NYISO, where buried infrastructure is more significant. This limitation restricted the study's ability to fully represent transmission networks in densely populated areas. This constraint was also observed by the low count of power lines in the OSM data as well. Additionally, while the dataset provides a static snapshot of component configurations, the lack of dynamic operational data limits its capacity to model network behavior during a geomagnetic event. Regarding methodology, there is an opportunity in the current application of voltage assignments, which were treated homogeneously in this study, to provide a more tailored perspective. Future research should focus on extracting more specific circuit breaker operating voltages according to attributes such as region, operator, or manufacturer information and rerunning the analysis to refine component assignments further.

This research supports the broader application of open-source mapping approaches to improve assessment models. By emphasizing collaborative data collection, the project enables the rapid development of detailed models, particularly relevant for studying geomagnetic disturbances and other specific risks to critical infrastructure. While this study prioritized isolating component data points that are considered most vulnerable to GICs, it also tracked components that may not directly relate to GICs but hold value for broader risk modeling. These efforts ensure that the dataset is versatile, providing opportunities for infrastructure operators to enhance system integrity.

\section*{Acknowledgments}\label{acknowledgements}
\addcontentsline{toc}{section}{Acknowledgments}

This material is based upon work supported by the NSF National Center for Atmospheric Research, which is a major facility sponsored by the U.S. National Science Foundation under Cooperative Agreement No. 1852977. The project upon which this article is based was funded through the NSF NCAR Early-Career Faculty Innovator Program under the same Cooperative Agreement. We also gratefully acknowledge funding from the ChronoStorm NSF RAPID grant (\#2434136), co-funded by the GEO/AGS Space Weather Research and the ENG/CMMI Humans, Disasters, and the Built Environment programs. We thank Liling Huang of the Department of Electrical and Computer Engineering at George Mason University for participating in the expert validation workshops.

\section*{Open Research}\label{open-research}
\addcontentsline{toc}{section}{Open Research}

While the annotation workflow and supporting software are openly documented, the raw component-level annotations are not publicly available because they link named infrastructure locations with detailed transformer counts, types, and substation configurations, creating critical infrastructure sensitivity concerns. Source code releases for the Coupled Space Weather Impact Model (C-SWIM) and the C-SWIM Mapping and Annotation Platform (C-MAP) are archived on Zenodo \cite{bor_2026_21987174,bor_2026_21987224}. The processed inputs and derived outputs supporting the Tennessee Valley Authority (TVA) and University of Illinois Urbana-Champaign 150-bus synthetic network (UIUC150) case studies are available in the accompanying Zenodo data and software archive \cite{bor_2026_21986729}. These materials include the geospatial network derived from OpenStreetMap (OSM) and Homeland Infrastructure Foundation-Level Data (HIFLD), along with the modeled GIC time series. The TVA GIC measurements used for validation are available in the May 2024 Geomagnetic Storm Data Repository \cite{wilkerson_2026_20090223}, and the observational methods are described by \citeA{wilkerson_gic--related_2026}.

\section*{Conflict of Interest}
The authors declare there are no conflicts of interest for this manuscript.

\newpage
\bibliography{repr_mapping_paper}

\FloatBarrier
\end{document}